\documentclass[journal]{IEEEtran}

\renewcommand{\baselinestretch}{0.97}

\ifCLASSINFOpdf
\else
   \usepackage[dvips]{graphicx}
\fi
\usepackage{url}

\usepackage{graphicx}
\usepackage{pdfpages}
\usepackage{setspace}
\usepackage{multirow}
\usepackage{amsmath}
\usepackage{epsfig}
\usepackage{array}
\usepackage{color}
\usepackage{subfigure}
\usepackage{xcolor}
\usepackage{amssymb}
\usepackage{url}
\usepackage{mathrsfs}
\usepackage{booktabs}
\usepackage{cite}
\usepackage{bbding}
\usepackage{upgreek}
\usepackage{amssymb}
\usepackage{amsfonts}
\usepackage[noend]{algpseudocode}
\usepackage{algorithmicx,algorithm}
\usepackage{amsmath,amsfonts,amssymb,graphicx}
\usepackage{mathrsfs}
\usepackage{flushend}

\begin{document}

\title{A Robust Maximum Likelihood Distortionless Response Beamformer based on a Complex Generalized Gaussian Distribution}

\author{Weixin Meng,
        Chengshi~Zheng,
        and Xiaodong~Li
	\thanks{The authors are with the Key Laboratory of Noise and Vibration Research, Institute of Acoustics, Chinese Academy of Science,
		Beijing, 100190, China, and also with University of Chinese Academy of Sciences, Beijing, 100049, China (email: cszheng@mail.ioa.ac.cn)}
	\thanks{Manuscript received February XX, 2021; revised XXXX XX, XX.}}

\markboth{Journal of \LaTeX\ Class Files, Vol. XX, No. XX, February 2021}
{Shell \MakeLowercase{\textit{et al.}}: Bare Demo of IEEEtran.cls for IEEE Journals}
\maketitle


\begin{abstract}
For multichannel speech enhancement, this letter derives a robust maximum likelihood distortionless response beamformer by modeling speech sparse priors with a complex generalized Gaussian distribution, where we refer to as the CGGD-MLDR beamformer. The proposed beamformer can be regarded as a generalization of the minimum power distortionless response beamformer and its improved variations. For narrowband applications, we also reveal that the proposed beamformer reduces to the minimum dispersion distortionless response beamformer, which has been derived with the ${{\ell}_{p}}$-norm minimization. The mechanisms of the proposed beamformer in improving the robustness are clearly pointed out and experimental results show its better performance in PESQ improvement.
\end{abstract}


\IEEEpeerreviewmaketitle

\section{Introduction}

\IEEEPARstart{M}{icrophone} array beamforming has been widely used to extract the desired speech and suppress both the interferences and the noise for speech communication and automatic speech recognition (ASR). Typically, there are two types of microphone array beamforming algorithms, where one is data-independent fixed beamformers and the other is data-dependent adaptive beamformers. Generally, the adaptive beamformers are more powerful in suppressing directional interferences adaptively than the fixed beamformers. There are many well-known adaptive beamformers including the minimum power distortions response (MPDR) beamformer \cite{mvdr,cox1973resolving,ehrenberg2010sensitivity}, the generalized sidelobe cancellation (GSC) \cite{griffiths1982alternative,gannot2004speech,talmon2009convolutive} and the multi-channel Wiener filtering (MWF) \cite{spatially-mwf,SDW-MWF,optimal-mwf,habets2011-mwf,wang2015gsc,zheng2018statistical,benesty2020iterative}. Both the MPDR beamformer and the GSC are quite sensitive to the steering vector error of the desired speech, while the performance of the MWF depends on the estimation accuracy of the second-order statistics of the desired speech and the interference-plus-noise component. Among these beamformers, the MPDR beamformer and its variations are the most widely studied and applied.



There are at least two ways to improve the robustness of the MPDR beamformer, where one is to improve the estimation accuracy of the steering vector of the desired speech \cite{cohen2004relative,markovich2009multichannel,Kellermann2013geometrically,pan2013performance,markovich2018performance,hu2020decoupled} and the other is to estimate the interference-plus-noise power spectral density (PSD) matrix to replace the noisy PSD matrix \cite{hendriks2011noise,ko2018limiting,gu2012robust,taseska2017nonstationary,koutrouvelis2019robust,pan2020estimation}. For practical applications, these two ways can be combined together to further improve the performance. Whereas one cannot expect that the steering vector can be estimated accurately and the interference-plus-noise PSD matrix does not contain any desired speech PSD matrix, especially in low signal-to-interference-plus-noise ratio (SINR) conditions. This paper focuses on improving the robustness of the MPDR beamformer without estimating the interference-plus-noise PSD matrix explicitly.


When assuming that the desired speech in the frequency domain follows a complex Gaussian distribution (CGD) with time-varying variances, a maximum likelihood distortionless response (MLDR) beamformer has been derived in \cite{MLDR} and it reduces the word error rates for ASR. When considering the signal, interferences and the noise are non-Gaussian distributed, a minimum dispersion distortionless response (MDDR) beamformer has been derived with the $\ell_p$-norm minimization for narrowband applications \cite{MDDR,robustMDDR2017}. The relationship between the MLDR beamformer and the MPDR beamformer has not been revealed clearly and its mechanism in improving the performance needs to be further clarified. Moreover, the best choice of $p$ in MDDR is not straightforward, which also needs to be studied in a more theoretical way.

In this letter, we derive a robust maximum likelihood distortionless response beamformer by introducing a complex generalized Gaussian distribution to model speech sparse priors \cite{2007-MMSE-CGGD,hendriks2009optimal}, which is refer to as the CGGD-MLDR beamformer. One can see that the proposed beamformer is a generalization of the MPDR beamformer and it can reduce to many existing variations of the MPDR beamformer. This letter also shows the mechanism of the CGGD-MLDR beamformer in improving the robustness of the MPDR beamformer. Finally, we propose an iterative optimization algorithm to obtained the optimal weight vector of the proposed beamformer. Experimental results show that the proposed beamformer can achieve better performance by properly choosing the shape parameter $p$.


The remainder of this letter is organized as follows. Section \uppercase\expandafter{\romannumeral2} presents problem formulation and related work. In section \uppercase\expandafter{\romannumeral3}, we derive the CGGD-MLDR beamformer and study its relationship with the MPDR beamformer and its variations, and the mechanism in improving the robustness is also presented. In section \uppercase\expandafter{\romannumeral4}, we study the performance of the CGGD-MLDR beamformer, and compare it with the MLDR and the MPDR beamformers. Section V gives some conclusions.

\section{Problem Formulation and Related Work}
We assume that a desired speech source and some independent directional noise sources impinge on an arbitrary shape microphone array consisting of $M$ microphones. By applying the short-time Fourier transform (STFT), the microphone signals can be written into vectors of $M$ so that $\mathbf{y}(k,l)={\left[ Y_{1}(k,l),...,Y_{M}(k,l)\right]^{T}}$, where $k$ denotes the frequency index and $l$ denotes the frame index, we have
\begin{equation}
\begin{aligned}
  \mathbf{y}(k,l)&={\mathbf{h}(k)}{S(k,l)}+\mathbf{v}(k,l) \\
 & ={\mathbf{x}(k,l)}+{\mathbf{v}(k,l)},
\end{aligned}
\end{equation}
where $S(k,l)$ denotes the desired speech; ${\mathbf{h}(k)}$ denotes the acoustic transform function (ATF) vector of the desired speech; $\mathbf{v}(k,l)$ denotes the interference-plus-noise vector.

The objective of a beamforming is to design a spatial filter $\mathbf{w}(k)$, which can be applied to extract the desired speech:
\begin{equation}
\begin{aligned}
	\label{S_estimation}
	{\widehat S}(k,l)&={\mathbf{w}^{H}(k)}{\mathbf{y}(k,l)}\\
	&={\mathbf{w}^{H}(k)}{\mathbf{h}(k)}S(k,l)+{\mathbf{w}^{H}(k)}{\mathbf{v}(k,l)}.
\end{aligned}
\end{equation}

The well-known MPDR beamformer aims to minimize the output power with the distortionless constraint on the desired direction, which can be given by:
\begin{equation}
	\label{MVDR_opi}
    \underset{\mathbf{w}(k)}{\mathop{\min}}E{\left\{ |{\mathbf{w}^{H}(k)}{\mathbf{y}{(k,l)}}|^{2}\right\}} \quad s.t. \quad {\mathbf{w}^{H}(k)\mathbf{h}(k)=1}.
\end{equation}
The close-formed solution of (\ref{MVDR_opi}) can be written as
\begin{equation}
	\label{w_mvdr}
  {{\mathbf{w}}_\text{MPDR}(k)}=\frac{{{ {\mathbf{R} }_\mathbf{yy}^{-1}(k) }}{{\mathbf{h}}(k)}}{\mathbf{h}^{H}(k){{{\mathbf{R} }_{\mathbf{yy}}^{-1}(k) }}{{\mathbf{h}}(k)}},
\end{equation}
where
\begin{equation}
	\begin{small}
	\begin{aligned}
	\mathbf{R}_{\mathbf{yy}}(k)&=E\left\{ \mathbf{y}(k,l)\mathbf{y}^{H}(k,l) \right\}  \\
	&={\lambda_{s}(k)}{\mathbf{{\Upsilon}}_{{ss}}(k)}+{\lambda_{v}(k)}{\mathbf{{\Upsilon}}_{\mathbf{vv}}(k)}\\
	\end{aligned}
\end{small}
\end{equation}
denotes the noisy PSD matrix; $\lambda_{s}(k)$ denotes the PSD of the desired speech; ${\mathbf{{\Upsilon}}_{{ss}}(k)}$ denotes the desired speech correlation matrix; ${\lambda_{v}(k)}$ denotes the PSD of the interference-plus-noise signal and ${\mathbf{{\Upsilon}}_{\mathbf{vv}}(k)}$ denotes the interference-plus-noise correlation matrix. In practice, the noisy PSD matrix needs to be replaced by using its sample covariance matrix, which can be given by:
\begin{equation}
	\begin{aligned}
		_\text{MPDR}{{\mathbf{\widehat{R}}}_{\mathbf{yy}}(k)}&=\sum\limits_{l=1}^{{{\mathcal{L}}}}{\mathbf{y}\left(k,l \right){{\mathbf{y}}^{H}}\left(k,l\right)}. \\
	\end{aligned}
\end{equation}

Accordingly, the estimated optimal weight vector of the MPDR beamformer can be expressed as
\begin{equation}
  	\label{w_mvdr_es}
  	{{\mathbf{\widehat w}}_{\text {MPDR}}(k)}=\frac{{{ \left(_\text{MPDR}{\mathbf{\widehat R} }_{\mathbf{yy}}^{-1}(k)\right) }}{{\mathbf{h}}(k)}}{\mathbf{h}^{H}(k){{  \left(_\text{MPDR}{\mathbf{\widehat R} }_{\mathbf{yy}}^{-1}(k)\right) }}{{\mathbf{h}}(k)}}.
\end{equation}

It is well-known that the MPDR beamformer is sensitive to the estimation error of the ATF error and the desired speech cancellation problem often occurs. To improve its robustness, the noisy PSD matrix should be replaced by the interference-plus-noise PSD matrix. For this purpose, one needs to distinguish the interference-plus-noise-only time-frequency bins from the noisy bins or the speech presence probability in each time-frequency bin needs to be estimated before updating the interference-plus-noise PSD matrix.

In \cite{MLDR}, the MLDR beamformer is derived and its optimal weight vector has the similar form as the MPDR beamformer, where the only difference is that the noisy PSD matrix in the MPDR beamformer is replaced by a weighted sample covariance matrix, which can be given by:
\begin{equation}
	\label{Ry_mldr}
_\text{CGD}\mathbf{{\widehat R}}_{\mathbf{yy}}(k)=\sum\limits_{l=1}^{\mathcal{L}}\frac{{{\bf y}\left( {k,l} \right){\bf y}^H \left( {k,l} \right)}}{{\lambda _{s} \left( {k,l} \right)}},
\end{equation}
where $\lambda _{s} \left( {k,l} \right) = E\left\{|S^2(k,l)|\right\}$ indicates the PSD of the desired speech in the $k$th bin of the $l$th frame. Note that $|S^2(k,l)|$ is often unknown prior and it needs to be replaced by its estimated value, i.e., $\widehat \lambda _{s} \left( {k,l} \right) = |\widehat S^2(k,l)|$ should be used for practical applications.

\section{Method}
\subsection{MLDR with a Complex Generalized Gaussian Distribution}
We assume that $S(k,l)$ follows a zero-mean complex generalized Gaussian distribution, where its probability density function can be expressed as:
\begin{equation}
 \label{pdf}
  \rho \left( S(k,l) \right)=\frac{p}{2\pi \gamma \Gamma \left( {2}/{p}\; \right)}{{e}^{-\frac{{{\left| S(k,l) \right|}^{p}}}{{{\gamma }^{{p}/{2}\;}}}}},  	
\end{equation}
where $\gamma>0$ is the scale parameter, $p$ is the shape parameter of this complex generalized Gaussian distribution, and $\Gamma \left( \cdot  \right)$ is the Gamma function. The complex generalized Gaussian can be divided into three groups including super-Gaussian, Gaussian, and sub-Gaussian, respectively. In this letter, the super-Gaussian distribution is considered for speech applications, i.e. $0<p<2$. In this case, $\rho \left( S(k,l) \right)$ can further be represented as a maximization over scaled Gaussian distributions with different variances, which can be expressed as:
\begin{equation}
	\label{convex_pdf}
	\rho \left( S(k,l) \right)={\underset{\lambda_s(k,l) >0}{\mathop{\max }}}{{\mathbb{N}}_{\mathbb{C}}}\left( S(k,l);0,\lambda_s(k,l)  \right)\psi \left( \lambda_s(k,l)  \right),
\end{equation}	
where ${\mathbb{N}}_{\mathbb{C}}(S(k,l);0,\lambda_s(k,l))$ denotes a complex Gaussian distribution with zero-mean and time-varying variances $\lambda_s(k,l)$; $\psi(\cdot)$ denotes a scaling function which is related to $\lambda_s(k,l)$. With this model, the weight vector $\mathbf{w}(k)$ should be optimized by maximizing the following likelihood function:
\begin{equation}
\begin{aligned}
& \max \prod\limits_{l=1}^{\mathcal{L}}{\underset{{{\lambda_s }(k,l)}>0}{\mathop{\max }}\,{{\mathbb{N}}_{\mathbb{C}}}\left( {S(k,l)};0,{{\lambda_s }(k,l)} \right)\psi \left( {{\lambda_s}(k,l)} \right)} \\
& s.t.\quad \mathbf{w}^{H}(k){{\mathbf{h}}(k)}=1 \\
\end{aligned}
\end{equation}
which is equivalent to minimize the negative log-likelihood with the weight vector $\mathbf{w}(k)$ and $\lambda_s(k,l)$, which can be expressed as:
\begin{equation}
\label{Min}
\begin{aligned}
& \underset{{{\lambda_s }(k,l)}>0}{\mathop{\min }}\,\sum\limits_{l=1}^{\mathcal{L}}{\left( \frac{{{\left| {{S}(k,l)} \right|}^{2}}}{{{\lambda_s }(k,l)}}+\log \left( \pi {{\lambda_s }(k,l)} \right)-\log  \psi \left( {{\lambda_s }(k,l)} \right) \right) } \\
& s.t.\quad \mathbf{w}^{H}(k){{\mathbf{h}}(k)}=1.
\end{aligned}
\end{equation}

It is worth noting that the optimization problem is correlated to the desired speech $S(k,l)$, which is unknown prior and it is the estimation target of the beamformer. We replace $S(k,l)$ with its estimation $\widehat S(k,l)$. A Lagrange multiplier method can be used to solve this optimization problem, and the cost function can be written as
\begin{equation}
	\label{lagrange}
	\mathcal{J}_{k}=\sum\limits_{l=1}^{\mathcal{L}}{ g({\mathbf{w}({k})},{\lambda_s(k,l)}) }+\alpha_{k}(\mathbf{w}^{H}(k){{\mathbf{h}}(k)}-1),
\end{equation}
where
\begin{equation}
\begin{small}
  \begin{aligned}
   g({\mathbf{w}(k)},{\lambda_s(k,l)})\!=\!\frac{{{\left| {{\widehat S}(k,l)} \right|}^{2}}}{{{\lambda_s }(k,l)}}+\log \left( \pi {{\lambda_s }(k,l)} \right)-\log  \psi \left( {{\lambda_s }(k,l)} \right),
    \end{aligned}
\end{small}
\end{equation}
and $\alpha_{k}$ denotes the Lagrange multiplier. The optimization problem need to optimize $\lambda_s(k,l)$ and $\mathbf{w}({k})$ simultaneously, which means that we cannot get a closed-form solution. In this letter, we propose an iterative optimization algorithm to solve this problem and the updating rule can be obtained by setting partial differentials
of the cost function with respect to the corresponding parameters at zero. Similar to \cite{Dereve-Sparce-2015}, by setting the partial differentials of $\mathcal{J}_{k}$ with respect to $\lambda_s(k,l)$ at zero, one can get
\begin{equation}
	\label{lambda_solution}
	{\lambda_s(k,l)}=\frac{2{\gamma }^{{p}/{2}}}{p}{{\left| {{\widehat S}(k,l)} \right|}^{2-p}},
\end{equation}
when $\lambda_s(k,l)$ is determined, $\mathbf{w}(k)$ is the solution of
\begin{equation}
	\label{wk_iteration}
	\underset{\lambda_s(k,l)>0}{\mathop{\min }}\,\sum\limits_{l=1}^{\mathcal{L}}{\frac{{{\left| \widehat S(k,l) \right|}^{2}}}{\lambda_s(k,l)}}\quad s.t.\quad \mathbf{w}^{H}(k)\mathbf{h}(k)=1.
\end{equation}

Finally, the solution of (\ref{wk_iteration}) can be given by
\begin{equation}
	\mathbf{\widehat w}_{\text{CGGD}}(k)=\frac{{{\left(_{\text{CGGD}}\mathbf{\widehat R}_{\mathbf{yy}}(k) \right)}^{-1}}{{\mathbf{h}}(k)}}{\mathbf{h}^{H}{{\left(_{\text{CGGD}}\mathbf{\widehat R}_{\mathbf{yy}}(k) \right)}^{-1}}{{\mathbf{h}}(k)}},
\end{equation}
where
\begin{equation}
	\begin{aligned}
	\label{Rx_iteration}
	_{\text{CGGD}}\mathbf{\widehat R}_{\mathbf{yy}}(k)=\sum\limits_{l=1}^\mathcal{L}{\frac{{{\mathbf{ y}}(k,l)}\mathbf{ y}^{H}(k,l)}{\lambda_s (k,l)}}
	=\sum\limits_{l=1}^\mathcal{L}{\frac{p{{\mathbf{ y}}(k,l)}\mathbf{ y}^{H}(k,l)}{2{\gamma }^{{p}/{2}}{\left| {{\widehat S}(k,l)} \right|}^{2-p}}}.
	\end{aligned}
\end{equation}

Because $\mathbf{\widehat w}_{\text{CGGD}}(k)$ is invariant to the constant scaling factor in (\ref{lambda_solution}), (\ref{Rx_iteration}) can be further reduced to
\begin{equation}
	\begin{aligned}
	\label{R_CGGD}
	_{\text{CGGD}} \mathbf{\widehat R}_{\mathbf{yy}}(k)=\sum\limits_{l=1}^{\mathcal{L}}{\frac{{{\mathbf{y}}(k,l)}\mathbf{y}^{H}(k,l)}{{\widehat \lambda_{s}^{(1-p/2)}(k,l)}}},
   \end{aligned}
\end{equation}
where $\widehat \lambda_{s}(k,l) = |\widehat S(k,l)|^2$ denotes the estimated PSD of the desired speech. To avoid dividing by zero, a small positive floor value $\delta \ge 0$ in the denominator should be introduced to improve the stability of the proposed beamformer.

In the iterative optimization algorithm, with the initialization $\mathbf{\widehat w}_{\text {CGGD}}^{0}(k)=\mathbf{\widehat w}_{\text{MPDR}}(k)$, we keep updating $\widehat \lambda_s(k,l)$ and $\mathbf{\widehat w}_{\text{CGGD}}(k)$ until reaching the maximum number of iterations. The whole algorithm is summarized in Algorithm 1.

\subsection{Related to the MPDR Beamfomer and its Variations}
This part discuss the relationship between the proposed CGGD-MLDR beamformer and the MPDR beamformer together with its improved variations. The proposed CGGD-MLDR beamformer is a generalized MPDR beamformer, where it can reduce to many existing improved versions of the MPDR beamformer for different values of the shape parameter $p$ in (\ref{pdf}). Obviously, we can have the following comments:

\begin{enumerate}
\item When $p=2$, the proposed CGGD-MLDR beamformer reduces to the well-known MPDR beamformer due to $_{\text{CGGD}}\mathbf{\widehat R}_{\mathbf{yy}}(k) \equiv \mathbf{\widehat R}_{\mathbf{yy}}(k)$ because of ${{\widehat \lambda_{s}^{(1-p/2)}(k,l)}} \equiv 1$.
\item When $p=0$, the proposed CGGD-MLDR beamformer becomes the newly proposed MLDR beamformer{\footnote{In {\cite{NTT-WPD-2}}, the MLDR was also called the weighted MPDR (wMPDR).}} {\cite{MLDR}} due to $_{\text{CGGD}}\mathbf{\widehat R}_{\mathbf{yy}}(k) \equiv _{\text{CGD}}\mathbf{\widehat R}_{\mathbf{yy}}(k)$ for $p=0$.
\item When $p$ is a positive value, for narrowband applications, the proposed CGGD-MLDR beamformer has the same form as the MDDR beamformer that is derived using the $\ell_p$-norm minimization {\cite{MDDR}}. Note that the proposed CGGD-MLDR beamformer gives clear guidelines on choosing the shape parameter $p$ in (\ref{pdf}), this is because it is derived from the maximum likelihood theory and $p$ relates to the distribution of the desired speech.
\end{enumerate}

\begin{algorithm}[t]
	\caption{CGGD-MLDR}
	\hspace*{0.02in} {\bf Input:}
	${\mathbf{y}(k,l)}$, $p$, and maximum iteration number $I$\\
	\hspace*{0.02in} {\bf Output:}
	$\mathbf{\widehat w}_{\text{CGGD}}^{I}(k)$ and $\widehat S^{I}(k,l)$
	\begin{algorithmic}[1]
		\State {\bf{Initialize:}} $\mathbf{\widehat w}_{\text{CGGD}}^{0}(k)={{\mathbf{\widehat w}}_{\text {MPDR}}}(k)$
		\For{$i=0,1,2,...,I{-1}$}
		\For {$l=1,2,...,\mathcal{L}$}
		\State Compute ${\widehat S}^{i}(k,l)={\mathbf(\mathbf{\widehat w}_{\text{CGGD}}^i(k))^{H}}{\mathbf{y}(k,l)}$
		\State Update ${\widehat \lambda_s^{i+1}(k,l)}={{\left| {{\widehat S}^{i}(k,l)} \right|}^{2-p}}$
		\State Compute
        \State$_{\text{CGGD}}\mathbf{\widehat R}_{\mathbf{yy}}^{i+1}(k)=_{\text{CGGD}}{\mathbf{\widehat R}_{\mathbf{yy}}^{i+1}(k)+{\frac{{{\mathbf{ y}}(k,l)}\mathbf{y}^{H}(k,l)}{\widehat \lambda_s ^{i+1}(k,l)}}}$
		\EndFor
		\State Update $	\mathbf{\widehat w}_{\text{CGGD}}^{i+1}(k)=\frac{{{\left(_{\text{CGGD}}\mathbf{\widehat R}_{\mathbf{yy}}^{i+1}(k) \right)}^{-1}}{{\mathbf{h}}(k)}}{\mathbf{h}^{H}(k){{\left( _{\text{CGGD}}\mathbf{\widehat R}_{\mathbf{yy}}^{i+1}(k) \right)}^{-1}}{{\mathbf{h}}(k)}}$
		\EndFor
		\State \Return $\mathbf{\widehat w}_{\text{CGGD}}^{I}(k)$ and $\widehat S^{I}(k,l)$
	\end{algorithmic}
\end{algorithm}

\begin{figure*}[t]
	\centerline{\includegraphics[width=1.8\columnwidth]{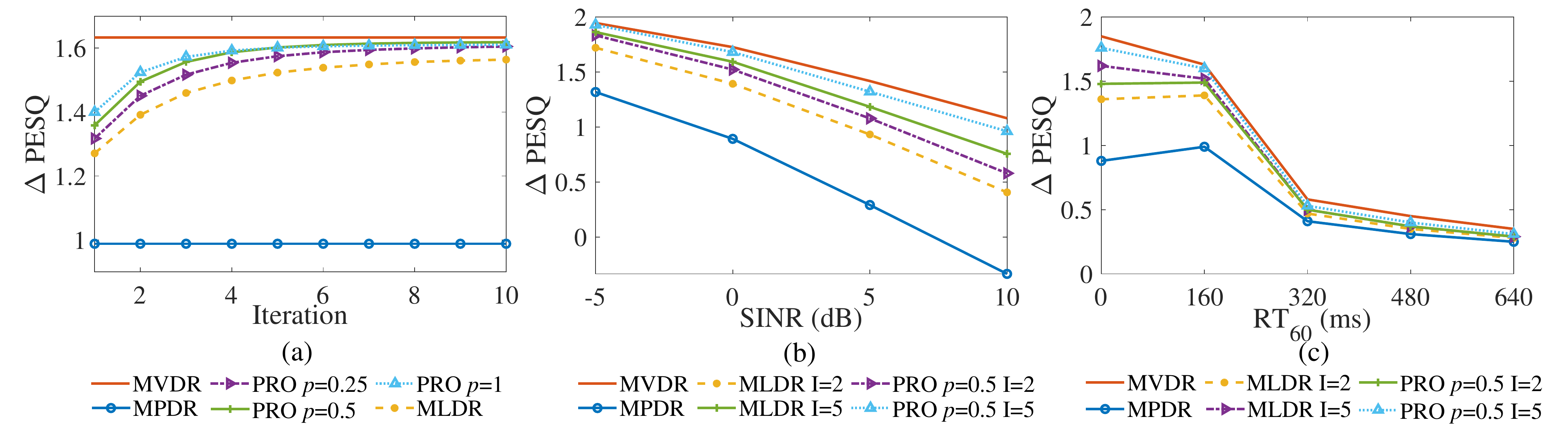}}
	\caption{Performance evaluation. (a) PESQ improvements versus the number of iterations. (b) PESQ improvements versus the input SINR. (c) PESQ improvements versus the reverberation time. "PRO" represents the proposed CGGD-MLDR beamformer for simplicity.}
	\label{Fig1}
\end{figure*}

\subsection{Mechanisms of the Proposed CGGD-MLDR Beamformer in Improving Robustness}
We assume that there are $\mathcal{L}_1$ interference-plus-noise-only frames among all $\mathcal{L}$ frames. And, we further assume that $\mathcal{L}$ and $\mathcal{L}_1$ is large enough so that the inter-correlation between ${\bf x}(k,l)$ and ${\bf v}(k,l)$ can be ignored and $\lambda _{s} \left( {k,l} \right)$ and $\lambda _{v} \left( {k,l} \right)$ do not change over time. Accordingly, (\ref{R_CGGD}) becomes
\begin{small}
\begin{equation}
\begin{aligned}
{}_{{\rm CGGD}}{\bf R}_{{\bf yy}} \left( k \right)& = \mathcal{L}_2\left( {\lambda _s \left( k \right)} \right)^{\frac{p}{2}} \mathbf{{\Upsilon}}_{ss} \left( k \right) \\ &+ \left( {\mathcal{L}_1\rho(k)  + \mathcal{L}_2\left( {\lambda _s \left( k \right)} \right)^{\frac{p}{2}} \frac{1}{{\varepsilon \left( k \right)}}} \right)\mathbf{{\Upsilon}}_{vv} \left( k \right),
\end{aligned}
\end{equation}
\end{small}
\noindent where $\rho(k) = {{\lambda _v \left( k \right)} \mathord{\left/
 {\vphantom {{\lambda _v \left( k \right)} \delta^{(1-p/2)} }} \right.
 \kern-\nulldelimiterspace} \delta^{(1-p/2)} }$ and $\varepsilon(k)=\lambda _s \left( k \right)/\lambda _v \left( k \right)$is the input SINR. $\mathcal{L} = \mathcal{L}_1 + \mathcal{L}_2$ with $\mathcal{L}_2$ the number of frames containing desired speech. One can see that ${}_{{\rm CGGD}}{\bf R}_{{\bf yy}} \left( k \right)$ is a linear combination of $\mathbf{{\Upsilon}}_{ss} \left( k \right)$ and $\mathbf{{\Upsilon}}_{vv} \left( k \right)$, we further define the ratio of the two combination coefficients, which is
 \begin{equation}
 r_p\left( k \right) = \frac{{\mathcal{L}_1 \rho \left( k \right) + \mathcal{L}_2 \left( {\lambda _s \left( k \right)} \right)^{\frac{p}{2}} \frac{1}{{\varepsilon \left( k \right)}}}}{{\mathcal{L}_2 \left( {\lambda _s \left( k \right)} \right)^{\frac{p}{2}} }}.
 \end{equation}

When $p=2$, we have $r_2\left( k \right) = {\mathcal{L} \mathord{\left/
 {\vphantom {\mathcal{L} {\left( {L_2 \varepsilon \left( k \right)} \right)}}} \right.
 \kern-\nulldelimiterspace} {\left( {\mathcal{L}_2 \varepsilon \left( k \right)} \right)}}$, and thus $r_2\left( k \right)$ is determined by the input SINR and the number of the desired speech frames among all $\mathcal{L}$ noisy frames. Note that the smaller $r_2\left( k \right)$ is, the more sensitive the MPDR beamformer is. When $p=0$, we have $r_0\left( k \right) = {{\left( {\mathcal{L}_1 \rho \left( k \right) + {{\mathcal{L}_2 } \mathord{\left/
 {\vphantom {{\mathcal{L}_2 } {\varepsilon \left( k \right)}}} \right.
 \kern-\nulldelimiterspace} {\varepsilon \left( k \right)}}} \right)} \mathord{\left/
 {\vphantom {{\left( {\mathcal{L}_1 \rho \left( k \right) + {{\mathcal{L}_2 } \mathord{\left/
 {\vphantom {{\mathcal{L}_2 } {\varepsilon \left( k \right)}}} \right.
 \kern-\nulldelimiterspace} {\varepsilon \left( k \right)}}} \right)} {\mathcal{L}_2 }}} \right.
 \kern-\nulldelimiterspace} {\mathcal{L}_2 }}$. Obviously, when $\rho(k)\varepsilon \left( k \right) \ge 1$, i.e., $\lambda_s(k) \ge \delta$, $r_0(k) \ge r_2(k)$ holds true always. This should be the reason that the MLDR beamformer can improve the robustness of the MDPR beamformer. For arbitrary values of $p \in [0\;2)$, one can easily derive that $r_p(k) \ge r_2(k)$ holds true if and only if $\lambda_s(k) \ge \delta$, which means that the CGGD-MLDR beamformer can be always more robust than the MPDR beamformer due to that $\delta$ is only a small positive value as mentioned above.

\section{Experiment Results}
This section evaluates the performance of the proposed CGGD-MLDR beamformer and compares it with the MPDR beamformer and the newly developed MLDR beamformer. The performance of the oracle minimum variance distortionless response (MVDR) beamformer is also presented to show the theoretical limit, where the interference-plus-noise PSD matrix is assumed to be known exactly. Ten 20s speech signals are taken from TIMIT corpus \cite{timit} and the babble noise is chosen form NOISEX-92 database \cite{noise92} and is split into multiple segments as interferences. The room impulse response is generated by using the image method \cite{habets2006room}, with a room of size $6m\times10m\times4m$. The reverberation time ranges from 0 to 640$ms$ with the interval 160$ms$. We consider a uniform linear array with $6$ microphones and $4cm$ inter-sensor distance which is placed at the center of the room. The desired speech is $2m$ away from the array center propagating from $\theta=0^{\circ}$, and two interferences propagate from $45^{\circ}$ and $-45^{\circ}$, respectively. In this evaluation, the PESQ improvement {\cite{PESQ}} is chosen as an objective measurement.
\subsection{Performance versus Iteration}
In the first experiment, we examine the performance of the beamformers versus the number of iterations with the input $\text{SINR=0dB}$ and the reverberation time 160$ms$. From Fig. {\ref{Fig1}} (a), one can see that the choice of $p$ seriously affects the performance of the CGGD-MLDR beamformer. Among them, it has the highest PESQ improvement with $p=0.5$ and it can gradually converge to the oracle MVDR, while MPDR have the lowest PESQ improvement. Moreover, one can see the CGGD-MLDR beamformer with $p=0.5$ provides much higher PESQ improvement using only very limited iterations, e.g., 2 to 3 iterations, which means that it has much faster convergence rate than the MLDR beamformer.
\subsection{Performance versus Input SINR}
In the following experiments, we only present the results of the proposed beamformer with $p=0.5$ and compare it with the MLDR and the MPDR beamformers for its best performance. Fig. {\ref{Fig1}} (b) plots the PESQ improvements versus the input SINR ranging from -5dB to 10dB with the reverberation time 160$ms$. We can see that the PESQ improvement reduces as the increase of the input SINR for all beamformers and the proposed CGGD-MLDR beamformer with $p=0.5$ is much better than the other beamformers. For the MPDR beamformer, the PESQ improvement can be negative due to that the target speech cancellation problem occurs in high input SINR conditions.
\subsection{Performance versus Reverberation Time}
In the third experiment, we study the performance of CGGD-MLDR with $\text{RT}_{60}={\left[0,160,320,480,640 \right]}ms$. It is shown in Fig. {\ref{Fig1}} (c) that, in low $\text{RT}_{60}$ scenarios, e.g 0 to 160$ms$, the proposed beamformer is significantly better than the MLDR beamformer, while with the increase of $\text{RT}_{60}$, the performance of all beamformers decreases and their performance becomes similar. When the reverberation time increases, it is well-known that the MPDR beamformer and its variations decrease their performance due to that the directional interferences have multiple strong reflections, which cannot be well suppressed by imposing nulls.

\section{Conclusion}
When a complex generalized Gaussian distribution is introduced to model the desired speech, we derive the CGGD-MLDR beamformer with the maximum likelihood criterion, which is a generalization of the MPDR and the MLDR beamformers. By properly choosing the shape parameter $p$, the proposed beamformer can achieve better performance than the MLDR beamformer in PESQ. The most attractive aspect is that the proposed beamformer with $p=0.5$ can converge in a much faster way than the MLDR beamformer, and thus the computational complexity can be decreased dramatically due to that the proposed beamformer needs much fewer iterations to achieve the same performance of the MLDR beamformer. Future work can concentrate on jointly optimizing denoising and dereverberation with the complex generalized Gaussian distribution of the desired speech.

\clearpage

\end{document}